\documentclass[fleqn]{article}
\usepackage{amsmath,amsfonts,graphicx}

\newcommand{\href}[2]{{#2}}
\newcommand{\pdfbookmark}[3][0]{\empty}

\title{Loop Model with Generalized Fugacity in Three Dimensions}
\author{
  Saburo Higuchi%
  \thanks{e-mail: hig@rice.c.u-tokyo.ac.jp}\\
  {\it Department of Pure and Applied Sciences,} \\
  {\it The University of Tokyo, Komaba} \\
  {\it  Komaba, Meguro, Tokyo 153-8902, Japan}
}
\date{1999/07/22, UT-KOMABA/99-9, cond-mat/9907335}

\renewcommand{\thefootnote}{\fnsymbol{footnote}}

\begin{document}

\begin{titlepage}
\setcounter{page}{0}
\thispagestyle{empty}
  \begin{flushright}
    July 1999\\
    \href{http://hep1.c.u-tokyo.ac.jp/local-preprints.html}{UT-KOMABA}/99-9\\
    \href{http://xxx.lanl.gov/abs/cond-mat/9907335}{cond-mat/9907335}
  \end{flushright}
\vspace*{\fill}

\begin{center}
  {
    \LARGE\bf
    \mbox{Loop Model with Generalized Fugacity} \\[0.5ex]
    \mbox{in Three Dimensions}
    } 
\vspace*{\fill}

\href{http://rice.c.u-tokyo.ac.jp/~hig/}{{\Large Saburo Higuchi}}%
\footnote{
  e-mail:
  \href{mailto:hig@rice.c.u-tokyo.ac.jp}{\texttt{hig@rice.c.u-tokyo.ac.jp}}
}\\[0.5ex]

\href{http://dbs.c.u-tokyo.ac.jp/labs/kisoka/}{\textit{Department of Pure and Applied Sciences,}} \\
\href{http://www.u-tokyo.ac.jp/}{\textit{The University of Tokyo}}, 
\href{http://www.c.u-tokyo.ac.jp/}{\textit{Komaba}}\\
\textit{Komaba, Meguro, Tokyo 153-8902, Japan}
  \end{center}
\vspace*{\fill}

\vspace*{\fill}
\begin{center}
{\large\bf Abstract}
\pdfbookmark[1]{Abstract}{Z}
\end{center}

A statistical model of loops on the three-dimensional lattice is
proposed and is investigated.
It is O($n$)-type  but has  loop fugacity that 
depends on global three-dimensional shapes of loops in a particular fashion.
It is shown that, despite this non-locality and the dimensionality,
a layer-to-layer transfer matrix can be constructed 
as a product of local vertex weights
for infinitely many points in the parameter space.
Using this transfer matrix, 
the site entropy is estimated numerically in the fully packed limit.
\smallskip
\vspace*{5ex}

\noindent
\href{http://www.aip.org/pacs/}{PACS}:
05.20.-y  
05.40.Fb  
05.50.+q  
36.20.-r  


\smallskip
\begin{verbatim}
rcsid: $Header: rfpl.tex,v 4.1 99/12/02 10:19:44 hig Exp $
\end{verbatim}
\end{titlepage}

\setcounter{footnote}{0}
\renewcommand{\thefootnote}{\arabic{footnote}}

\section{Introduction} \label{sec:introduction}
Loop models are interesting examples of statistical models of
extended objects. They are related to  
the O($n$) spin model \cite{Nienhuis:exact,DoMuNiSc:duality}, 
a surface growth model \cite{KoHe::four-coloring}, 
the self-avoiding walk \cite{DuSa:exact},
the protein folding problem \cite{ChDi:protein}, 
and so on. It includes the fully packed loop model \cite{BlNi:honeycomb}
and the Hamiltonian
cycle problem \cite{OrItDo:hamiltonian,Suzuki:regular,Higuchi:aspectratio}
as particular limits.

The partition function of an O($n$) loop model on a lattice
with $N$ sites at the inverse temperature $x$ is given by 
\begin{equation}
  Z_{\text{loop}}(n,x^{-1}) = \sum_{c\in\mathcal{C}}
    x^{\mathcal{N}_\mathrm{S}(c)-N} n^{\mathcal{N}_\mathrm{L}(c)}
    \quad \quad(n,x\in\mathbb{R}).
  \label{loop-partition}
\end{equation}
The summation is taken over the set $\mathcal{C}$ 
of all the non-intersecting loop 
configurations drawn along links of the lattice.
The number of loops and that of sites visited by them 
are denoted by 
$\mathcal{N}_\mathrm{L}(c)$ and $\mathcal{N}_\mathrm{S}(c)$, respectively.

One may hope to study the model \eqref{loop-partition} by the transfer 
matrix approach. For $n\in\mathbb{Z}_+$, 
this is done in a simple way; one introduces 
link variables whose values are either occupied states with one of $n$
colors or an unoccupied state and lets them interact on sites.
A transfer matrix is written 
as a product of vertex weights straightforwardly.

For  $n\not\in\mathbb{Z}_+$, however,
the partition sum \eqref{loop-partition} cannot be rewritten in terms
of local degrees of freedom such as link variables in a simple way.
It is not trivial to have a \emph{local} transfer matrix%
\footnote{
  The use of the connectivity basis is discussed in subsection
  \protect\ref{sec:connectivity}.
  }.
I say a transfer matrix is \emph{local} when its component is written
as a product of weights each of which 
is determined by the local state configuration around a lattice site.

It is surprising that, in two dimensions,  
$n\not\in\mathbb{Z}_+$ models 
admit a mapping onto a state sum model with a local vertex weight and 
thus have local transfer matrices 
\cite{Baxter:coloring,Nienhuis:CoulombGas,BaBlNiYu:packedloop}.
In fact, by choosing $s\in\mathbb{C}$ satisfying $n=s+s^{-1}$,
$Z_{\text{loop}}(n,x^{-1})$ can be written as
\begin{equation}
\!\!\!\!\!\!
Z_{\text{loop}}(n,x^{-1})
=\sum_{c\in\mathcal{C}} x^{\mathcal{N}_\mathrm{S}(c)-N} 
                              (s+s^{-1})^{\mathcal{N}_\mathrm{L}(c)}
= \sum_{c\in\overline{\mathcal{C}}}  x^{\mathcal{N}_\mathrm{S}(c)-N} 
\prod_{L\in\overline{\mathcal{L}}(c)} s^{\pm1},
\label{directed}
\end{equation}
where $\overline{\mathcal{C}}$ is the set of loop configurations
with a direction associated with each loop.
The set $\overline{\mathcal{L}}(c)$ consists 
of all the directed loops in a configuration $c$.
A loop with the (counter-)clockwise direction is given a weight 
$s^{+1}$ $(s^{-1})$.
This weight can be realized by associating 
$s^{+1/4}$($s^{-1/4}$) with each right(left)-turn site
and the model can be regarded as a state sum model with a local vertex 
weight.
This trick has made the study of two-dimensional loop models very fruitful.

Physics of loops in three dimensions is very attractive.
It is  realistic in the context of condensed matter physics.
There has been a continuous suspicion that two-dimensional ones
have missed some important ingredient in real physics, 
\textsl{e.g.} the protein folding problem.
Three-dimensional loops have also rich mathematical structures.
For instance, loops can be knotted or linked 
in three dimensions\cite{Nechaev:entangled}.
It is noted that a number of attractive proposals 
have been made to generalize the loop model to higher dimensions
\cite{WiKa:geometric,WiKa:coloredmanifolds}.

The analysis of loop models and their generalizations in higher
dimensions is, however,  
extremely hard to perform. 
Needless to say, the number of configurations increases considerably.
For fugacity $n\not\in\mathbb{Z}_+$, which includes the interesting
case of the self-avoiding walk ($n=0$),
no way of constructing local transfer matrices is known.
This is because specialties of two dimensions cannot be used  
to simplify problems any more.
The mapping \eqref{directed} makes use of the fact that the
a directed loop in two dimensions turns around just once either clockwisely or 
counter-clockwisely.
It appears that this kind of tricks  never works in
higher dimensions. 

In this article, I propose a model
which generalizes \eqref{loop-partition} in a fashion
specific to three dimensions.
It is furnished with loop fugacity that depends on the global
three-dimensional shape of loops.  
I show that, despite this generalization which makes 
the model even more non-local,
a local transfer matrix for the system  can be constructed 
for a number of choices of fugacity.
These choices include the ones that give zero or non-integer weight to 
loops.

This article is organized as follows.
In section \ref{sec:fugacity}, I define a loop model in three
dimensions generalizing \eqref{loop-partition}.
Its local transfer matrix is constructed for a family of points in the 
parameter space in section \ref{sec:transfer_matrix}.
In section \ref{sec:entropy},
this transfer matrix is numerically diagonalized to yield an estimate
of the site entropy in the fully packed limit $x^{-1}=0$.
In the last section \ref{sec:discussions}, I discuss my results and 
their relation to combinatorial problems.
In an appendix, a technical issue on the block-diagonalization of the
transfer matrix is addressed.

\section{Generalized fugacity}\label{sec:fugacity}
I define a statistical model of loops 
on the three-dimensional simple cubic lattice 
$\mathbb{Z}^3=
\{\sum_{i=1}^3 m_i \mathbf{e}_i\in\mathbb{R}^3|m_i\in\mathbb{Z}\}$ ,
$\mathbf{e}_i \cdot\mathbf{e}_j=\delta_{ij}$.
The partition function is given by
\begin{equation}
  Z_{\text{loop}}[n](x^{-1})
= \sum_{c\in\mathcal{C}}
x^{\mathcal{N}_\mathrm{S}(c)-N}
\prod_{L\in\mathcal{L}(c)} n(A(L)),
\label{rloop-partition}
\end{equation}
where $\mathcal{L}(c)$ is the set of loops in a configuration $c$.
The loop fugacity $n$ is now promoted to a function 
which 
depends on the shape of $L\in\mathcal{L}(c)$ 
through a quantity $A(L)\in\mathbb{R}$ defined below.

To define $A(L)$, 
one begins with associating  
a closed trajectory on the unit sphere with each loop $L$.
One picks a direction for $L$.
On every point $\mathbf{x}\in L \subset \mathbb{R}^3$ except 
for sites where $L$ makes a turn,
there is a unit tangential vector $\mathbf{v}(\mathbf{x})$ to $L$;
it is either of $\pm \mathbf{e}_i, \;i=1,2,3$.
One may regard $\mathbf{v}(\mathbf{x})$
as a mapping from $L\setminus \text{(`turn-sites')}$ 
to the unit sphere $\mathrm{S}^2$.

\begin{figure}[tb]
  \begin{center}
    \leavevmode
    \includegraphics[scale=0.5]{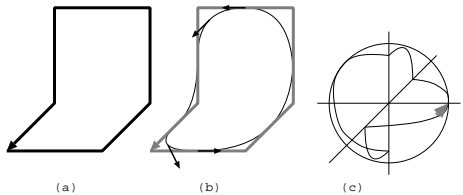}
    \caption{
      The definition of $A(L)$.
      (a) The original loop $L$ with a direction 
      associated. (b) The loop smoothed at turn-sites.
      (c) Trajectory of the tangent vector 
      on the unit sphere.
      An area of $A(L)=2\pi$ is enclosed.
      \label{fig:smoothing}
      }
  \end{center}
\end{figure}

As one walks along $L$,
$\mathbf{v}(\mathbf{x})$  jumps from a point to another on $\mathrm{S}^2$.
One can naturally interpolate these points to 
define a continuous trajectory $\mathbf{v}: L\rightarrow \mathrm{S}^2$.
One has only to declare that $\mathbf{v}(\mathbf{x})$ moves 
along the geodesic (of length $\tfrac12\pi$) on $\mathrm{S}^2$
at each turn-site.
This is equivalent with smoothing a loop in neighborhoods of turn-sites
keeping it within the plane (Fig.\ref{fig:smoothing}).
Then one defines $A(L)$ to be 
the oriented area encircled in the right of the trajectory
$\mathbf{v}(\mathbf{x})$.
On the lattice $\mathbb{Z}^3$, $A(L)$ takes values
\begin{equation}
  A(L)=\frac12m\pi \; (m\in\mathbb{Z}) \label{A-unit}.
\end{equation}
In two dimensions, the quantity $A(L)$ takes values $\pm2\pi$
and this signature corresponds to that of $s^{\pm1}$ in \eqref{directed}.
Therefore \eqref{rloop-partition} incorporates an essential
ingredient of three-dimensional loops and is regarded 
as a natural generalization of \eqref{directed}.

As is evident from the above construction, there is certain ambiguity
for the value of $A(L)$.
First, because the trajectory is drawn on a closed surface of area $4\pi$,
$A(L)$ is well-defined up to $4\pi$.
Second, the signature of $A(L)$ is changed when the picked 
direction of $L$ is reversed.
I require that  $n(\cdot)$ in \eqref{rloop-partition} absorbs this ambiguity.
Hence, it should satisfy
\begin{gather}
  n(A)=n(A+4\pi)     \label{A-modfourpi},\\
  n(-A)=n(A)         \label{A-signature}.
\end{gather}
Eqs. \eqref{A-unit}, \eqref{A-modfourpi},\eqref{A-signature} 
imply that the fugacity function $n(\cdot)$ 
can be specified
by five parameters $n(A), A=0,\frac12\pi,\pi, \frac32\pi,2\pi$.

In spite of the above restriction on $n(A)$, the model \eqref{rloop-partition}
includes many interesting cases. Consider, for example, fugacity
\begin{equation}
  n(A)=n_0 \delta^{(4)}_0(A),   \label{delta-n}
\end{equation}
with $n_0\in\mathbb{R}$ and 
\begin{equation}
  \delta^{(a)}_b(A)=
  \begin{cases}
    1 & \text{if\ }  \left(\frac{A}{\pi}-b\right)\equiv0 \bmod a,\\
    0 & \text{otherwise}.
  \end{cases} 
  \label{periodic_delta}
\end{equation}
The sum in \eqref{rloop-partition} is then restricted  
to configurations which consist only of loops 
with the oriented area $A \equiv 0 \mod 4\pi$.
It should be interesting to compare the site entropy with that of the
model with  $n(A)=n_0$.
It is also tempting to ask whether such an additional constraint
changes the critical behavior or not. 
The present case reminds one of 
the fully packed loop model in two dimensions.
Its universality class differs from that of densely packed
loop phase when 
the additional constraint that the loop length 
must be even is imposed \cite{BaBlNiYu:packedloop,Jacobsen:universality}.

\section{Transfer matrices from local vertex weights}%
\label{sec:transfer_matrix}
In order to construct a local layer-to-layer transfer matrix for
the loop model \eqref{rloop-partition}, 
I define a vertex model and show that it is equivalent with
\eqref{rloop-partition}. 

\begin{figure}[tb]
  \begin{center}
    \leavevmode
    \begin{center}
    \includegraphics[scale=0.5]{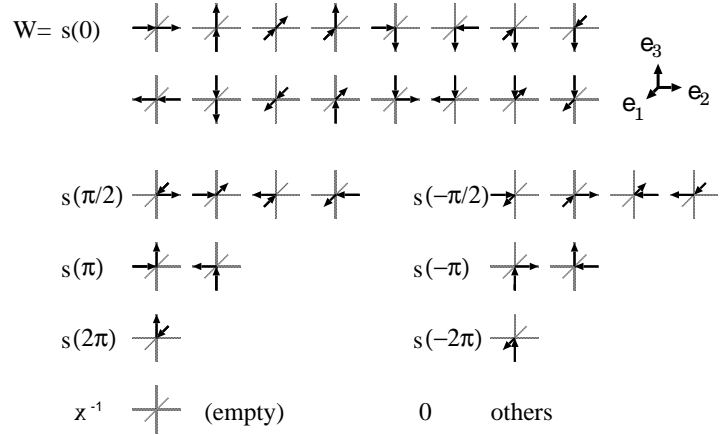}      
    \end{center}
    \caption{Vertex weights $W$ for $Z_{\text{vertex}}[s](x^{-1})$.
      \label{fig:vertex_weight}
      }
  \end{center}
\end{figure}

The local degree of freedom $z$ of the vertex model lives on each link 
$\langle \mathbf{r}, \mathbf{r}\pm\mathbf{e}_i\rangle$, 
$\mathbf{r}\in\mathbb{Z}^3$.
It takes one of three values $\leftarrow$, $\rightarrow$,  and  $-$(empty).
On each site,  six neighboring link variables interact 
by the vertex weight $W$ defined in Fig. \ref{fig:vertex_weight},
where $s(\omega)$ is a function that satisfies
\begin{equation}
 s(A_1)\times s(A_2)=s(A_1+A_2)
 \label{s-representation}
\end{equation}
and is specified further below.
The partition function of the vertex model is
\begin{equation}
  Z_{\text{vertex}}[s](x^{-1})
=\sum_{z=\leftarrow, \rightarrow,-} 
\;\;
\prod_{\mathbf{r}\in\mathbb{Z}^3} 
W( \{z(\langle \mathbf{r}, \mathbf{r}\pm\mathbf{e}_i\rangle)\}).
\label{vertex-partition}
\end{equation}
Evidently, the partition function \eqref{vertex-partition} has a
local transfer matrix which is a product of $W$'s.

Now I  show that \eqref{vertex-partition} for an appropriate $s$
is equivalent with \eqref{rloop-partition}.
Because the weight $W$ is nonzero only when there is one incoming and
one outgoing arrow, contribution to the partition sum
\eqref{vertex-partition} comes only from 
the set $\overline{\mathcal{C}}$ of directed loop configurations
\begin{equation}
  Z_{\text{vertex}}[s](x^{-1})
  =\sum_{c\in\overline{\mathcal{C}}} x^{\mathcal{N}_\mathrm{S}(c)-N} 
  \prod_{L\in\overline{\mathcal{L}}(c)} 
\left[ \;\;
 \prod_{\mathbf{r}\in L \cap \mathbb{Z}^3} s(\omega(\mathbf{r}))
\right]
\label{vertex-dloop-partition}
\end{equation}
where $s(\omega(\mathbf{r}))$ is the weight for  the vertex 
at $\mathbf{r}\in\mathbb{Z}^3$ in Fig.\ref{fig:vertex_weight}.

The factor in the square bracket in \eqref{vertex-dloop-partition}
is associated with a directed loop component $L$.
To evaluate this quantity, it is crucial to observe that
\begin{equation}
A(L)=\sum_{\mathbf{r}\in L\cap\mathbb{Z}^3} \omega(\mathbf{r}).
\label{area-local}
\end{equation}
Actually, the weight system in Fig.\ref{fig:vertex_weight} is designed 
to have this property
in refs.\cite{AsItKo:randomwalk,AsItKo:diffusion,Itoi:smooth}
in the context of random walk with a spin factor %
\cite{Polyakov:transmutation,IsItMu:geometric}.
Combining the properties \eqref{area-local} and \eqref{s-representation},
one finds that the factor in the square bracket is simply $s(A(L))$.

Using the same trick as that used in \eqref{directed}, one can further
write the partition function \eqref{vertex-dloop-partition}
as a sum over undirected loop configurations
\begin{equation}
  Z_{\text{vertex}}[s](x^{-1})
=\sum_{c\in\mathcal{C}} x^{\mathcal{N}_\mathrm{S}(c)-N} 
\prod_{L\in\mathcal{L}(c)} ( s(A(L))+s(-A(L))).
\end{equation}
Therefore, 
if $n(A)$ in the loop model  \eqref{rloop-partition} can be written as
\begin{equation}
 n(A)=s(A)+s(-A)
 \label{fugacity-decomposition}
\end{equation}
with a function $s$ that satisfies \eqref{s-representation},
then 
the vertex model partition function $Z_{\text{vertex}}[s](x^{-1})$ 
defined above is equal to \eqref{rloop-partition}.

The requirement \eqref{s-representation} together with 
the restriction 
\eqref{A-unit}, 
\eqref{A-signature} and  \eqref{A-modfourpi} 
on $n(A)$ forces $s(A)$ to have a simple form
\begin{equation}
  s(A)= e^{iJ A}
\end{equation}
with $J\in\mathbb{Z}/2$.
It is enough \cite{AsItKo:randomwalk} to consider the cases $J=0,
\tfrac12, 1, \tfrac32$ and $2$ because of \eqref{A-unit}.
Hereafter, one introduces a shorthand notation
\begin{equation}
  Z_J(x^{-1}) = 
Z_{\text{loop}}[ n(A)= e^{iJ A}+e^{-iJ A} ](x^{-1}).
\end{equation}

The vertex weights at $J=0$ and $2$ enjoy a special property
$s(\omega)=s(-\omega)$.
This enables one to define a model with 
only two microscopic states $\leftrightarrow$ and $-$:
\begin{align}
  Z'_{\text{vertex}}[s](x^{-1})
=&\sum_{z=\leftrightarrow, -} 
\prod_{\mathbf{r}\in\mathbb{Z}^3} 
W( \{z(\langle \mathbf{r}, \mathbf{r}\pm\mathbf{e}_i\rangle)\})\nonumber\\
=&\sum_{c\in\mathcal{C}} x^{\mathcal{N}_\mathrm{S}(c)-N} 
\prod_{L\in\mathcal{L}(c)} s(A(L)),
\end{align}
which I denote by $Z_{0'}$ and $Z_{2'}$.

The fugacity functions corresponding to 
$J=0, \tfrac12, 1, \tfrac32,2,0'$ and $2'$ are listed in Table
\ref{tab:generators}. 
They indeed give $n(A)\le0$ or $n(A)\not\in\mathbb{Q}$ for some loops.
\begin{table}[tb]
  \begin{center}
    \leavevmode
    \begin{tabular}{|l|l|}
      \hline
      $J$ &   $n(\cdot)$          \\
      \hline\hline
      ${0}$        &  $2\delta^{(\tfrac12)}_0(\equiv2)$   \\
      ${\tfrac12}$  & 
$2(\delta^{(4)}_0-\delta^{(4)}_2)
+\sqrt{2}
(\delta^{(4)}_{\tfrac12}
+\delta^{(4)}_{\tfrac72}
-\delta^{(4)}_{\tfrac32}
-\delta^{(4)}_{\tfrac52})
 $\\
      $1$           & $2(\delta^{(2)}_0 - \delta^{(2)}_1)$ \\
      ${\tfrac32}$  & 
$2(\delta^{(4)}_0-\delta^{(4)}_2)
-\sqrt{2}
(\delta^{(4)}_{\tfrac12}
+\delta^{(4)}_{\tfrac72}
-\delta^{(4)}_{\tfrac32}
-\delta^{(4)}_{\tfrac52})
 $ \\
      ${2}$        &  $2(\delta^{(1)}_0 - \delta^{(1)}_{\tfrac12})$  \\
\hline
      ${0'}$        &  $\delta^{(\tfrac12)}_0(\equiv1)$   \\
      ${2'}$        &  $\delta^{(1)}_0 - \delta^{(1)}_{\tfrac12}$ \\
\hline
    \end{tabular} 
  \end{center}
  \caption{ The generalized fugacity $n$ 
    which generates the semigroup of allowed models.
    The function   $\delta^{(a)}_b(\cdot)$  is defined in 
    eq.\protect\eqref{periodic_delta}.
    \label{tab:generators}
    }
\end{table}

Although only finite number of vertex models
$J=0,\tfrac12,1,\tfrac32,2,0'$ and $2'$ have been constructed  above, 
it is possible  to construct an infinite number of ones by taking
the direct sum of the space of their microscopic states.
More precisely,
one generalizes the link variable to 
take one of $2q+1$ $(q\in\mathbb{Z}_+)$ states:
$\leftarrow_k, \rightarrow_k$ with the $k$ labeling colors
$k=1,\ldots,q$ and an uncolored empty state $-$.
Introducing parameters 
  $J_k\in\{0,\tfrac12,1,\tfrac32,2\}$,
the vertex weight assignments in Fig.\ref{fig:vertex_weight}
are supplemented by additional rules:
\begin{itemize}
\item If the both two arrows have the $k$-th color,
  then  $W=e^{i J_k \omega}$.
\item If the two colors do not agree, $W=0$.
\end{itemize}
The cases $J_k=0', 2'$ are handled in the obvious manner.

The fugacity of the 
`direct sum' model $Z_{J_1\oplus J_2 \oplus \cdots \oplus J_q}$
is simply the sum:
\begin{gather}
n(A) = \sum_{k=1}^q \left[
 (e^{iJ_k A} + e^{-iJ_k A})\times
 B(J_k)\right],\\
B(J)=
\begin{cases}
  1       & ( J=0,\frac12,1,\frac32,2), \\
  \frac12 & ( J=0',2').
\end{cases}
\end{gather}
One immediately notices  that 
\begin{equation}
  Z_{0'\oplus0'} = Z_{0}, \quad\quad  Z_{2'\oplus2'} = Z_{2}.
\end{equation}
Thus the fugacity functions expressible via  vertex models
form an infinite semigroup under addition
\footnote{
  This direct sum operation may be used for the lattice
  construction 
  \protect
  \cite{AsItKo:randomwalk,AsItKo:diffusion,Itoi:smooth}
  of higher-spin three-dimensional field theories.
}
generated by
$J=0',\tfrac12,1,\tfrac32$ and $2'$.

One can also take the `tensor product' of the space of microscopic states of 
$Z_{J_1}$  and $Z_{J_2}$ to define a model $Z_{J_1\otimes J_2}$.
Let the link variable take five values 
$(z_1,z_2)=
  \uparrow\uparrow,   \uparrow\downarrow,  
  \downarrow\uparrow, \downarrow\downarrow,$ and $||$.
The vertex weight $W$ is defined to be the product of $W$'s with
$J=J_1$ and $J_2$.
Then the loop fugacity becomes
\begin{equation}
n(A) = 
\left[
 (e^{iJ_1 A} + e^{-iJ_1 A})\times
 B(J_1)\right] \times 
\left[
 (e^{iJ_2 A} + e^{-iJ_2 A})\times
 B(J_2)\right].
\end{equation}
However, a new fugacity function cannot be realized because 
the partition function $Z_{J_1\otimes J_2}$ is equivalent with 
an appropriate direct sum
\begin{equation}
  Z_{J_1\otimes J_2} = Z_{\overline{J}_1\oplus\cdots\oplus\overline{J}_q}
\end{equation}
corresponding to the decomposition rule of the representation of SU(2).

\section{Entropy estimates}\label{sec:entropy}
I numerically diagonalize the transfer matrices constructed in section 
\ref{sec:transfer_matrix}.
Throughout this section, I concentrate on the fully packed limit $x^{-1}=0$
where all the sites are visited by a loop.
This simple case is in fact very interesting case;
in two dimensions, this limit yields a new universality class with a
shifted central charge 
on several bipartite lattices and has been attracting much attention
\cite{Jacobsen:universality,%
  Higuchi:decorated,%
  DiGuKr:fully_packed_eulerian,%
  GuKrNi:hamiltonian_eulerian}.
It would be interesting to look at the limit 
where the two strong constraints are combined:
the fully-packing constraint $x^{-1}=0$ and the constraint
\eqref{delta-n} on the shape of loops.

The site entropy in the thermodynamic limit is defined by 
\begin{equation}
  f[n](\infty) 
= \lim_{N\rightarrow\infty} \frac{1}{N} \log Z_{\text{loop}}[n](x^{-1}=0).
\label{thermodynamic-entropy}
\end{equation}
I evaluate this quantity on 
quasi-one-dimensional geometry $L_1\times L_2 \times L_3,
L_3\rightarrow\infty $ while $L_1,L_2$ kept finite 
by  calculating the largest eigenvalue of the transfer
matrix in an appropriate sector.

Let $T$ be the layer-to-layer transfer matrix in
$+\mathbf{e}_3$ (vertical) 
direction for the vertex model defined in \eqref{vertex-partition}.
Then $T$ acts on linear combinations of 
arrays of $L_1\times L_2$ vertical (colored) arrows.
One can take either hard-wall or periodic boundary condition 
in the horizontal directions.

It is important to note that the transfer matrix $T$ commutes
with the operator giving the net flow of arrows of $k$-th color in $+\mathbf{e}_3$ direction
\begin{equation}
 d_k = (\#\uparrow_k) - (\#\downarrow_k), \label{baxter}
\end{equation}
which is understood as
\begin{equation}
   d_k = (\#\updownarrow_k) \bmod 2
\end{equation}
for $J_k=0'$ and $2'$.
Thus  $T$ is block-diagonalized as 
\begin{equation}
  T=\bigoplus_{\mathbf{d}} T_{\mathbf{d}},
  \quad \mathbf{d}=(d_1,\ldots,d_q).
\label{block-decomposition}
\end{equation}

The quantity \eqref{thermodynamic-entropy} is obtained as 
\begin{equation}
  f[n](\infty) = 
\lim_{L_1, L_2\rightarrow\infty} 
\frac{1}{L_1L_2} \log |\lambda_{\mathbf{0}}^{0}(L_1,L_2)|,
\end{equation}
where $\lambda^i_{\mathbf{d}}(L_1,L_2)$ is the $i$-th largest 
eigenvalue of $T_{\mathbf{d}}(L_1,L_2)$. 
The condition $\mathbf{d}=\mathbf{0}$ excludes unwanted configurations
that have unbalanced arrows traveling along the infinite direction.
Shown in Tables \ref{tab:irreducible} and \ref{tab:reducible}
are the finite-$L_1,L_2$ results%
\footnote{I have also measured several leading  eigenvalues of 
  $T_\mathbf{d}$ with $\mathbf{d}=(d_1,\ldots,d_q)$, $d_k=0,\pm1$.
  These are related to correlation length of operators in the
  theory.
  These results will be reported elsewhere.
}.
The asymmetric Lanczos algorithm is utilized for the present sparse
eigenproblem. 
\begin{table}[tb]
  \begin{center}
    \leavevmode
    \begin{tabular}{|l|l|l|l|l|l|l|}
        \hline 
$J$ 
      & $2\times2$(h) & $3\times3$(h) & $3\times4$(h) 
      & $2\times2$(p) & $3\times3$(p) & $3\times4$(p) \\
      \hline\hline
      ${0}$        & 0.54202495 & 0.59145447 & 0.63524092 
                   & 1.0585126 & 0.83841678 & 0.83340128  \\
      ${\tfrac12}$ & 0.27123680 & 0.33576248 & 0.35050951 
                   & 0.79451346 & 0.55900063 & 0.55895924  \\
      $1$           & 0.51585927 & 0.50234791 & 0.55646223 
                    & 0.97170402 & 0.69812631 & 0.69832061 \\
      ${\tfrac32}$  & 0.35592318 & 0.35908194 & 0.37136801 
                    & 0.79451346 & 0.57435935 & 0.49496387 \\
      ${2}$        & 0.49499647 & 0.45468972 & 0.51615498 
                   & 1.0406166 & 0.66200716 &  0.67440981 \\
\hline
      ${0'}$        & 0.46298939 & 0.55650697 & 0.60072954  
                    & 0.91847381 & 0.79631788 & 0.80135760 \\
      ${2'}$        & 0.38697370 & 0.37695844 & 0.42272584 
                    & 0.87898824 & 0.55608904 & 0.58497412 \\
\hline
\hline
     & $4\times4$(p) &\multicolumn{5}{|l}{}\\
\cline{1-2}\cline{1-2}
$0'$ & 0.81947983 &\multicolumn{5}{|l}{}\\
$2'$ & 0.60931946 &\multicolumn{5}{|l}{}\\
\cline{1-2}
    \end{tabular}
  \end{center}
  \caption{
    The site entropy estimated numerically.
    $L_1\times L_2$ is the size of a layer while (p) and (h) mean 
    periodic and hard-wall boundary conditions in a layer.
    \label{tab:irreducible}
    }
\end{table}

\begin{table}[tb]
  \begin{center}
    \begin{tabular}{|l|l|l|l|l|l|} \hline
    $\bigoplus_k J_k$  & $n$ & $2\times2$(h) & $3\times3$(h) 
                             & $2\times2$(p) & $3\times3$(p) \\
\hline\hline
      ${0'\oplus2'}$ & $2\delta^{(1)}_0$ & 0.52330515 & 0.57390934 
      &  1.0502400 & 0.81503626 \\
      ${0\oplus0}$ & 4 &  0.64498133 & 0.65020710 
                       & 1.2354255 & 0.90747958 \\
      ${0\oplus2}$ & $4\delta^{(1)}_0$  & 0.63331428 & 0.62765833 
                   & 1.232054 & 0.88008432 \\
      ${0'\oplus2'\oplus1}$ & $4 \delta^{(2)}_0$ & 0.63216104 & 0.61664018 
                            & 1.2110124 & 0.86121744 \\
      ${0\oplus0\oplus0\oplus0}$ & $8$ & 0.76911076 &
                                 & 1.4501153 & 1.0108338   \\
      ${0\oplus0\oplus2\oplus2}$ & $8\delta^{(1)}_0$ & 1.3280134 &
                                 &  1.4490048 & 0.98982172  \\
      ${0\oplus1\oplus1\oplus2}$ & $8\delta^{(2)}_0$& 0.75985245 &
                                 & 1.4364447  & 0.97947213  \\
      ${0'\oplus2'\oplus\frac12\oplus1\oplus\frac32}$ & $8\delta^{(4)}_0$ & 0.36464934 &
            & 1.3863922  &  0.88604978 \\
      ${\frac12\oplus\frac32}$ &$4(\delta^{(4)}_0-\delta^{(4)}_2)$ 
& 0.47382047 & 0.41770663 
& 1.0397208 &    0.63630800 \\
      ${0\oplus\frac12}$ & $\not\in\mathbb{Q}$  & 0.40013109 & 0.54670471 
             & 1.1686609 & 0.84652598 \\
      ${0\oplus\frac32}$ & $\not\in\mathbb{Q}$ & 0.44457252 & 0.55636026 
 & 1.1686609 & 0.84694306 \\
      ${0'\oplus0\oplus\frac12}$ & $\not\in\mathbb{Q}$ 
    & 0.48094569 & 0.57807813 
    & 1.2520659 & 0.62795992  \\
     \hline
\hline
\cline{1-4}
        &       & $3\times4$(h) &$3\times4$(p) & \multicolumn{2}{|l}{}\\
\cline{1-4}
$0'\oplus2'$ & $2\delta^{(1)}_0$ 
& 0.61935581 & 0.81580656 &\multicolumn{2}{|l}{}\\
\cline{1-4}
    \end{tabular} 
  \end{center}

  \caption{
    The site entropy estimated numerically.
    $L_1\times L_2$ is the size of a layer while (p) and (h) mean 
    periodic and hard-wall boundary conditions in a layer.
    \label{tab:reducible}
    }
\end{table}
The obstacle in making $L_1$ and $L_2$ large in the actual
numerical work is of course the exponential growth of the
dimensionality of the transfer matrix.
The selection of $\mathbf{d}=\mathbf{0}$ sector helps 
to reduce the dimensionality, though the improvement is polynomial.
For a direct sum $Z_{J_1 \oplus \cdots \oplus J_q}$,
the dimensionality is estimated to be
\begin{equation}
\dim (\mathbf{d}=\mathbf{0} \text{ sector})
\sim
(1 + 2u + p )^{L_1L_2}\times 2^{-p} 
\left(\frac{3}{4\pi L_1L_2}\right)^{u/2},
\end{equation}
where $p=\#\{k| J_k=0' \text{\ or } J_k=2'\}$, $u=q-p$.
One notices that it is better to recast $Z_{2'\oplus 2'}$ as $Z_2$
if one is interested in its $\mathbf{d}=\mathbf{0}$ sector.

The exponential growth is severe even after 
restricting to  the $\mathbf{d}=\mathbf{0}$ sector.
In order to increase $L_1L_2$ 
as much as possible within the available computer resources, 
I have further decomposed $T_\mathbf{0}$ with respect to the 
eigenvalue of shift (lattice momentum) operator 
for the periodic boundary case where the translational symmetry is present.
I have looked at the zero-momentum sector ${T_\mathbf{0}}^{(0,0)}$ 
as described in the appendix \ref{sec:zero_momentum}.
It is quite natural to expect that the largest eigenvalue lies there.
By this decomposition, the dimensionality of the eigenproblem 
is reduced, at most,  by $(L_1 L_2)^{-1}$.
As a drawback, 
the matrix ${T_\mathbf{0}}^{(0,0)}$ becomes less sparse than 
the original $T_\mathbf{0}$.
With both effects combined, some improvements in the memory
usage and the CPU time are observed.
Thus the analysis of larger systems becomes possible for the  periodic
boundary case, as seen in Tables \ref{tab:irreducible} and \ref{tab:reducible}.

\section{Discussions} \label{sec:discussions}
In the comparison between the periodic and the hard-wall boundary
conditions, 
one notices that the periodic case has always larger site entropy.
This is because many of the loops that wind  non-trivially in the
horizontal directions satisfy $A(L)=0$ and the loops with $A(L)=0$
contribute to every partition sum with a positive fugacity.

The numerical works in the present study have been carried out on
modest workstations.  Unfortunately, information in the thermodynamic
limit $L_1,L_2\rightarrow\infty$ is out of reach in the present
analysis.
For the study of criticality, ref.\cite{AsItKo:randomwalk},
where random walks with the weight in Fig.\ref{fig:vertex_weight} are studied,
is quite suggestive. It is reported that
Euclidean symmetry is not always recovered even in the continuum limit.

I discuss relations with combinatorial problems below.

\subsection{Even and odd number of loops of a specific type}
For most allowed values of $J$,
the loop fugacity takes both positive and negative values.
Some interesting combinatorial information is encoded in 
these models.
For example, the linear combinations
$\tfrac12( Z_{0'} \pm Z_{2'} )$ counts the number
of loop configurations
such that there are 
even (odd) number of loops for which  $\frac{2}{\pi}A(L)\equiv1\mod2$,
\textsl{e.g.},
\begin{equation}
  \frac12( Z_{0'} \pm Z_{2'} ) =
  \begin{cases}
   \sum_{c\in\mathcal{C}_{\text{even},\frac12}} 1,\\
   \sum_{c\in\mathcal{C}_{\text{odd},\frac12}} 1,    
  \end{cases}
  \label{evenodd}
\end{equation}
where
$ \mathcal{C}_{\text{even(odd)},a}$ is the subset of $\mathcal{C}$ 
and is defined by the following properties.
\begin{itemize}
\item $c\in \mathcal{C}_{\text{even(odd)},a}$  contains even (odd)
  number of loops with $A(L)/a\pi \equiv 1 \mod 2$.
\item All other loops in $c\in \mathcal{C}_{\text{even(odd)},a}$ 
  satisfy  $A(L)/a\pi \equiv 0 \mod 2$.
\end{itemize}
Similarly,
the quantities 
$  \frac12( Z_{0'\oplus2'} \pm Z_{1} )$ and 
$  \frac12({Z_{0'\oplus2'\oplus1} \pm Z_{\frac12\oplus\frac32}})$
are interpreted as the sums over
$\mathcal{C}_{\text{even(odd)},1}$
and
$\mathcal{C}_{\text{even(odd)},2}$.

Of the two $Z$'s in \eqref{evenodd},
the one with the larger leading eigenvalue
dominates the sum in the limit $L_3\rightarrow\infty$ 
studied in section \ref{sec:transfer_matrix}.
In finite geometries, both terms contribute to yield an exact number.

\subsection{Self-avoiding walk}
The partition function \eqref{loop-partition} in 
the limit $n\rightarrow0$ 
corresponds to the enumeration of self-avoiding walks.
Self-avoiding walks in three-dimensions have mainly been studied by
the exact enumeration method due to the lack of transfer matrix
formalism as pointed out in the introduction.

The model I propose in this paper can be regarded
as a step forward to overcome this difficulty;
in the models $Z_{0'\oplus2'}, Z_{0'\oplus2'\oplus1}$ and  
$Z_{0'\oplus2'\oplus\frac12\oplus1\oplus\frac32}$, 
the fugacity is set zero for families of loops.
This is, however, achieved by paying the cost of 
having larger $n$ for another family of loops.
Within the present construction, loops with $A(L)\equiv0\mod 4\pi$
cannot have weights different from the number of possible link states.
Thus the partition function listed above serve only as a very loose
upper bound for the entropy of self-avoiding walks.

The problem of construction of a local transfer matrix to enumerate
self-avoiding walks on three-dimensional lattices
still remains open.

\subsection{Mapping to ribbon configurations}  
\begin{figure}[tb]
  \begin{center}
    \leavevmode
    \begin{center}
    \includegraphics[scale=0.8]{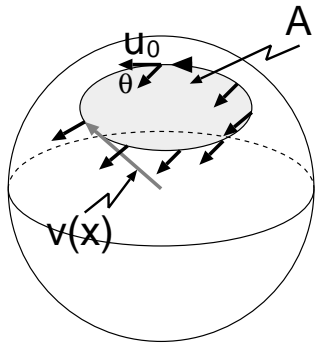}      
    \end{center}
    \caption{When $\mathbf{u}_0$, the tangent vector at
      $\mathbf{v}(\mathbf{x}_0)$, is parallel-transported 
      along the trajectory on $\mathrm{S}^2$,
      it receives holonomy whose magnitude (angle $\theta$) is equal to the
      oriented area $A(L)$ the path encloses.
      \label{fig:holonomy}}
  \end{center}
\end{figure}

The oriented area defined in \eqref{rloop-partition}
has a nice geometric interpretation as holonomy.
The tangent vector $\mathbf{v}(\mathbf{x})$ 
moves along a trajectory on $\mathrm{S}^2$.
Let the unit vector tangential to
this trajectory at $\mathbf{v}(\mathbf{x_0})$ be $\mathbf{u_0}$
(Fig.\ref{fig:holonomy}).
Consider the parallel transport (in the sense of riemannian
geometry) of $\mathbf{u}_0$
along the trajectory $\mathbf{v}(\mathbf{x})$ on $\mathrm{S}^2$.

\begin{figure}[tb]
  \begin{center}
    \leavevmode
    \begin{center}
    \includegraphics[scale=0.8]{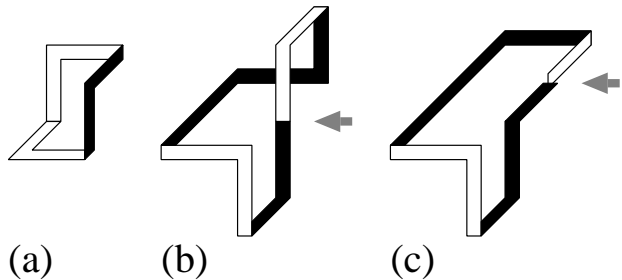}      
    \end{center}
    \caption{
      Examples of allowed ways of bending ribbons at sites 
      (except for the ones indicated by gray arrows) 
      The holonomy is accumulated at the sites indicated  by gray arrows.
      The right and the reverse sides of ribbons are represented by 
      white and black colors.
      \label{fig:ribbon}
      }
  \end{center}
\end{figure}

When $\mathbf{u}_0$ is transported back to 
$\mathbf{v}(\mathbf{x}_0)$, it gains some holonomy 
( the angle $\theta$ in Fig.\ref{fig:holonomy}).
This holonomy angle is given by the integration of the scalar curvature of
$\mathrm{S}^2$ over the domain encircled by the trajectory and is
nothing but the oriented area $A(L)$.
In the real space $\mathbb{Z}^3\subset\mathbb{R}^3$, 
the holonomy described above is nicely kept track of 
by broadening the loop segment to a `ribbon' with the
distinction of the right and the reversed sides.
The parallel transportation can be recasted 
as a rule of bending ribbons on sites, which is 
shown  in Fig.\ref{fig:ribbon}.
Among the loop configurations shown in Fig.\ref{fig:ribbon},
the partition sum $Z_{\text{loop}}$ with fugacity
$n=\delta^{(2)}_0$, $\delta^{(1)}_0$ and $\delta^{(\frac12)}_0$
receives contribution from
$\{\text{(a)}\}$,
$\{\text{(a),(b)}\}$,
and
$\{\text{(a),(b),(c)}\}$, respectively.

For $n=4\delta^{(2)}_0$, the sum in \eqref{rloop-partition} is
over ribbon loop configurations without mismatch.
In that interpretation,  the coefficient four is naturally regardes
as the number of directions the right side can face.
Therefore, $Z_{\text{loop}}[n=4\delta^{(2)}_0](x^{-1})$ is nothing but the
generating function of the number of allowed ribbon configurations%
\footnote{%
  This observation clearly indicate a simple way of 
  constructing transfer matrices for this  fugacity function.
  The link variable represents the ribbon which can be placed in four ways.
  The rule of bending is implemented in the vertex weight.
  The construction in section \ref{sec:transfer_matrix}, however, has
  an advantage; the size of the matrix can be reduced more by the choice of the
  sector $\mathbf{d}=\mathbf{0}$. 
}.

Similarly, the partition sum for $n=2\delta^{(1)}_0$ can be
interpreted as the sum over the configurations of ribbons without the
distinction between the right and the reverse sides, 
while in the case $n=\delta^{(\tfrac12)}_0$, the loop segment is just a
chord.
This interpretation suggests that 
\eqref{rloop-partition} may be regarded as a model of polymers with 
various partially broken axial symmetry by, for instance, the presence of
side chains. 

\subsection{Comparison with the connectivity basis}
\label{sec:connectivity}
The connectivity basis\cite{ScHiKl:compact,BlNi:square} is very powerful 
in that one can always write a transfer matrix for a loop model with
respect to it.
It has been very useful for numerical calculation in two-dimensions.

Nevertheless, I have avoided the use of the connectivity basis 
in the present work. The reason is the following.
First, its fundamental degrees of freedom are not the link variables
and the transfer matrix with respect to it is not local.
Local transfer matrices have  merits even in two dimensions.
Namely, it paved the way to the Bethe ansatz
solution\cite{Baxter:exactly,BaBl:honeycomb} 
and the conformal field theoretic description 
\cite{Cardy:geometrical,JaKo:fieldtheory}
via Coulomb gas representation.
Second, in three dimensions and higher, the size of connectivity
basis grows considerably because of the lack of the planarity constraint.
It is not clear if it is effective to perform numerical calculation in this 
basis.
In two dimensions, the present basis is as good as the connectivity one
\cite{BaBlNiYu:packedloop,Higuchi:decorated}.

I suppose it is very important to see how useful the connectivity basis in
three dimensions is and to try to improve the efficiency of the
calculation in that basis.

\section*{Acknowledgments}
\pdfbookmark{Acknowledgements}
I thank S. Tanimura for useful discussions on geometric phases.
I gratefully acknowledge useful conversations with 
M.~Asano, E.~Guitter, C.~Itoi,  S.~Hikami, K.~Minakuchi, and J.~Suzuki.
I thank T.~Iwai and Y.~Uwano for discussions and for hospitality at
Kyoto University, where a part of this work was done.

This work was supported by the Ministry of Education, Science
and Culture under Grant 08454106 and 
\href{http://rice.c.u-tokyo.ac.jp/~hig/kaken.html}{10740108} 
and by \href{http://www.jst.go.jp}{Japan Science and Technology Corporation}
under 
\href{http://www2.jst.go.jp/crest/}{CREST}.  \medskip

\appendix
\section{Projection to zero momentum subspace}
\label{sec:zero_momentum}
  In this short note, I describe the block-diagonalization of the
  transfer matrix 
  with respect to the eigenvalues of the lattice momentum operators.
  The zero-momentum subspace and 
  a reduced transfer matrix which acts on it 
  are explicitly constructed\footnote{%
    It may well improve the efficiency just to 
    choose a zero-momentum state as the initial Lanczos vector in the sparse
    algorithm without explicitly constructing a transfer matrix in the 
    subspace as is done in the text.
    }.

  One may start with the $(2q+1)^{L_1L_2}$-dimensional whole space of colored
  arrow configurations or an eigenspace of the operator $\mathbf{d}$.
  One considers the matrix elements in the basis $u_i, (i=1,\ldots,m)$,
  each of which represents a single arrow configuration
  such as $\uparrow\downarrow|\cdots \uparrow|\downarrow$:
  \begin{equation}
    T u_i = \sum_{j=1}^m T^j_i u_j.
  \end{equation}
  In this natural basis, the matrix $T$ becomes sparse.

  Let $S_1$ and $S_2$ be  discrete shift operators
  in the horizontal directions.
  Then the vectors 
  \begin{equation}
    v_j = \sum_{a=0}^{L_1-1}\sum_{b=0}^{L_2-1}
                               (S_1)^a
                               (S_2)^b u_j
  \end{equation}
  are zero-momentum ones.

  One classifies the index set as $\{1,\ldots,m\}=\sqcup_{I=1}^{M}
  V_I$ by an equivalence relation  $i\sim j \Leftrightarrow v_i=v_j$.
  Then $I=1,\ldots,M$ labels the zero-momentum subspace.
  The $(I,J)$-component of the block matrix is simply
\begin{equation}
  {({T_\mathbf{0}}^{(0,0)})_I}^J= 
\sum_{j \in V_J} {T_i}^j      \quad (i\in V_I).
\end{equation}
This procedure is fairy easy to implement in the sparse algorithm.

Evidently, a slight modification of the above procedure 
enables one to focus on a chosen non-zero momentum subspace.
It will be useful for identifying excited states.

It is noted that the above block-decomposition can be applied 
even if the seam factor is present, 
\emph{e.g.} to two-dimensional O($n$) model with cylinder topology.
One can make the system translationally
invariant by distributing the seam factor among 
all horizontal links.
I have checked that this prescription improves the efficiency of the
the enumeration of Hamiltonian cycles 
perfomed in ref.\cite{Higuchi:decorated}
although the weight system becomes system size dependent.

\pdfbookmark[1]{References}{X}

\bibliographystyle{apply}
\bibliography{mrabbrev,shrtjour,polymer,rwalk,textbook,mypubl}
\end{document}